\documentclass[twocolumn,showpacs,preprintnumbers,amsmath,amssymb]{revtex4}

\usepackage{graphicx}
\usepackage{dcolumn}
\usepackage{bm}
\usepackage[caption=false]{subfig}
\usepackage{color}
\usepackage[backref=page]{hyperref}
\usepackage{graphicx}
\usepackage{rotating}

\usepackage{amsfonts}
\usepackage{amssymb}
\usepackage{amsmath}
\usepackage{upgreek}

\hypersetup{
        colorlinks=true,          
      linkcolor=blue,          
    citecolor=blue,        
    urlcolor=green,           
}

\begin{document}


\title{Universal Loss Dynamics in a Unitary Bose Gas}

\author{Ulrich~Eismann$^{1,3}$}%
\altaffiliation{Present Address: Toptica Photonics AG, Lochhamer Schlag 19, 82166 Gr\"afelfing, Germany}
\thanks{These authors contributed equally to this work.}
\author{Lev~Khaykovich$^{1,2}$}
\thanks{These authors contributed equally to this work.}
\author{S\'ebastien~Laurent$^1$}
\author{Igor~Ferrier-Barbut$^1$}
\altaffiliation{Present Address: 5. Physikalisches Institut and Center for Integrated Quantum Science and Technology,
Universit\"at Stuttgart, Pfaffenwaldring 57, 70550 Stuttgart, Germany}
\author{Benno~S.~Rem$^1$}
\altaffiliation{Present Address: Institut f\"ur Laserphysik, Universität Hamburg, Luruper Chaussee 149, Building 69, D-22761 Hamburg, Germany}
\author{Andrew~T.~Grier$^1$}
\altaffiliation{Present address: Department of Physics, Columbia University, 538 West 120th Street, New York, NY 10027-5255, USA}
\author{Marion~Delehaye$^1$}
\author{Fr\'ed\'eric~Chevy$^1$}
\author{Christophe~Salomon$^1$}
\author{Li-Chung~Ha$^3$}
\author{Cheng~Chin$^3$}

\affiliation{$^1$Laboratoire Kastler Brossel, ENS-PSL Research University, CNRS,  UPMC, Coll\`ege de France, 24 rue Lhomond, 75005, Paris, France}
\affiliation{$^2$Department of Physics, Bar-Ilan University, Ramat-Gan 52900, Israel}
\affiliation{$^3$James Franck Institute, Enrico Fermi Institute and Department of Physics,  University of Chicago, Chicago, IL 60637, USA}


\date{\today}

\begin{abstract}
The low temperature unitary Bose gas is a fundamental paradigm in few-body and many-body physics, attracting wide theoretical and experimental interest. Here we first present a theoretical model that describes the dynamic competition between two-body evaporation and three-body recombination in a harmonically trapped unitary atomic gas above the condensation temperature. We identify a universal magic trap depth where, within some parameter range, evaporative cooling is balanced by recombination heating and the gas temperature stays constant. Our model is developed for the usual three-dimensional evaporation regime as well as the 2D evaporation case. Experiments performed with unitary $^{133}$Cs and $^7$Li atoms fully support our predictions and enable  quantitative measurements of the 3-body recombination rate in the low temperature domain. In particular,  we measure for the first time the Efimov inelasticity parameter $\eta_\ast = 0.098(7)$ for the 47.8-G $d$-wave Fesh\-bach resonance in $^{133}$Cs. Combined $^{133}$Cs and $^7$Li experimental data allow investigations of loss dynamics over two orders of magnitude in temperature and four orders of magnitude in three-body loss. We confirm the  $1/T^2$  temperature universality law up to the constant $\eta_\ast$.
\end{abstract}

\maketitle


\section{Introduction}
Resonantly interacting Bose systems realized in ultracold atomic gases are attracting growing attention thanks to being among the most fundamental systems in nature and also among the least studied.
Recent theoretical studies have included hypothetical BEC-BCS type transitions~\cite{Y-Li04,Romans04,Radzihovsky08,Koetsier09,Cooper10} and, at unitarity, calculations of the universal constant connecting the total energy of the system with the only energy scale left when the scattering length diverges: $E_n=\hbar^2n^{2/3}/m 	$~\cite{Cowell02,Song09,Y-Lee10,Diederix11}.
The latter assumption itself remains a hypothesis as the Efimov effect might break the continuous scaling invariance  of the unitary Bose gas and introduce another relevant energy scale to the problem.
A rich phase diagram of the hypothetical unitary Bose gas at finite temperature has also been predicted~\cite{Li12,Piatecki14}.

In experiments, several advances in the study of the resonantly interacting Bose gas have recently been made using the tunability of the s-wave scattering length $a$ near a Feshbach resonance.
The JILA group showed signatures of beyond-mean-field effects in two-photon Bragg spectroscopy performed on a $^{85}$Rb BEC~\cite{Papp08}, and the ENS group quantitatively studied the beyond mean-field Lee-Huang-Yang corrections to the ground state energy of the Bose-Einstein condensate~\cite{Navon11}.
Logarithmic behavior of a strongly interacting 2D superfluid was also reported by the Chicago group~\cite{Ha2013}.
Experiments have also started to probe the regime of unitarity ($1/a=0$ directly.
Three-body recombination rates in the non-degenerate regime have been measured in two different species, $^7$Li~\cite{Rem13} and $^{39}$K~\cite{Fletcher13}, and clarified the temperature dependence of the unitary Bose gas lifetime.
In another experiment, fast and non-adiabatic projection of the BEC on the regime of unitarity revealed the establishment of thermal quasi-equilibrium on a time scale faster than inelastic losses~\cite{Makotyn13}.

In a three-body recombination process three atoms collide and form a dimer, the binding energy of which is transferred into kinetic energies of the colliding partners.
The binding energy is usually larger than the trap depth and thus leads to the loss of all three atoms. Because three-body recombination occurs more frequently at the center of the trap, this process is  associated with ``anti-evaporative'' heating (loss of atoms with small potential energy) which competes with two-body evaporation and leads to a non trivial time dependence for the sample temperature.
In this paper, we develop a theoretical model that describes these atom loss dynamics. We simultaneously take into account two and three-body losses to quantitatively determine  each of these contributions.
We predict the existence of a magic value for the trap-depth-over-temperature ratio where residual evaporation compensates for three-body loss heating and maintains the gas temperature constant within some range of parameters.
We then apply our model to analyze the loss dynamics of $^{133}$Cs and $^{7}$Li unitary Bose gases prepared at various temperatures and atom numbers.
Comparing  measurements in these two different atomic species  we find the dynamics to be universal, i.e. in both systems the three-body loss rate is found to scale universally with temperature. Excellent agreement between theory and experiment confirms that the dynamic evolution of the unitary Bose gas above the condensation temperature can be well modelled by the combination of two and three-body interaction processes.

\section{Model}
\label{sec:Model}
A former study developed for measuring three-body decay in trapped  $^{133}$Cs~\cite{Weber03} atoms has proposed a model to describe the time evolution of the atom number $N$ and the temperature $T$ taking into account the three-body recombination induced loss and the heating associated with it.
This model is valid in the limit of deep trapping potentials (trapping depth much larger than the atom's temperature) and for temperature independent losses. Here we generalize this model to include evaporation induced cooling and the associated atom loss, as well as the temperature dependence of the three-body loss rate.

\subsection{Rate equation for atom number}
The locally defined three-body recombination rate $L_3 n^3(r)/3$ leads, through integration over the whole volume, to the loss rate of atoms:
\begin{equation}
\frac{{\rm d}N}{{\rm d}t}=-3\int{\frac{L_3 n^3(r)}{3} {\rm d}^3 r}=-L_3 \langle n^2 \rangle N,
\label{Eq_RateN3b}
\end{equation}
where the factor of $3$ in front on the integral reflects the fact that all $3$ atoms are lost per each recombination event.
In the following, we neglect single-atom losses due to collisions with the background gas and we assume that  two-body inelastic collisions are forbidden, a condition which is fulfilled for atoms polarized in the absolute ground state.

An expression for the three-body recombination loss coefficient at unitarity for a non-degenerate gas has been developed in Ref.~\cite{Rem13}. Averaged over the thermal distribution it reads:
\begin{eqnarray}
L_3&=&\frac{72\sqrt{3}\pi^2\hbar\left(1-e^{-4\eta_*}\right)}{mk_{\rm th}^6}
\nonumber\\
&&\times\int_0^\infty \frac{\left(1-|s_{11}|^2\right)e^{-k^2/k_{\rm th}^2}k\,{\rm d}k}{|1+(kR_0)^{-2is_0}e^{-2\eta_*}s_{11}|^2},
\label{Eq_FullL3}
\end{eqnarray}
where $k_{\rm th}=\sqrt{mk_{\rm B} T}/\hbar$, $R_0$ is the three-body parameter, and the Efimov inelasticity parameter $\eta_*$ characterizes the strength of the short range inelastic processes. Here, $\hbar$ is the reduced Planck's constant, $k_{\rm B}$ is the Boltzmann's constant, and $s_0=1.00624$ for three identical bosons\,\cite{Efimov1970}.
The matrix element $s_{11}$ relates the incoming to outgoing wave amplitudes in the Efimov scattering channel  and shows the emerging discrete scaling symmetry in the problem (see for example Ref.~\cite{Tung14}). Details are given in
the supplementary material to Ref.~\cite{Rem13} for the calculation of $s_{11}(ka)$, where $a$ is the scattering length and $k$ is the relative wavenumber of the colliding partners.
Because of its numerically small value for three identical bosons at unitarity, we can set $|s_{11}|=0$ and $L_3$ is well approximated by:

\begin{equation}
L_3\approx\frac{\hbar^5}{m^3}36\sqrt{3}\pi^2\frac{1-e^{-4\eta_*}}{(k_{\rm B} T)^2}=\frac{\lambda_3}{T^2},
\label{Eq_ApproxL3}
\end{equation}
where $\lambda_3$ is a temperature-independent constant. Assuming a harmonic trapping potential, we directly express the average square density $\langle n^2\rangle$ through $N$ and $T$. In combination with Eq.~(\ref{Eq_ApproxL3}), Eq.~(\ref{Eq_RateN3b}) is represented as:
\begin{equation}
\frac{{\rm d}N}{{\rm d}t}=-\gamma_3 \frac{N^3}{T^5},
\label{Eq_RateN3bSimple}
\end{equation}
where
\begin{equation}
\gamma_3 =\lambda_3\left(\frac{m\overline{\omega}^2}{2\sqrt{3}\pi k_{\rm B}}\right)^3,
\label{Eq_Gamma3}
\end{equation}
with $\overline{\omega}$ being the geometric mean of the angular frequencies in the trap.

To model the loss of atoms induced by evaporation, we consider time evolution of the phase-space density distribution of a classical gas:
\begin{equation}
f(\textbf{r},\textbf{p})=\frac{n_0 {\lambda_{\mathrm{dB}}}^3}{(2\pi\hbar)^3}  e^{-U(\textbf{r})/k_{\rm B}T}e^{-p^2/2mk_{\rm B}T},
\label{Eq_PhaseSpaceDist}
\end{equation}
which obeys the Boltzmann equation. Here $n_0$ is the central peak density of atoms, $\lambda_{\mathrm{dB}}=(2\pi \hbar^2/mk_{\rm B}T)^{1/2}$ is the thermal de Broglie wavelength,
and $U(\textbf{r})$ is the external trapping potential. The normalization constant is fixed by the total number of atoms, such that $\int f(\textbf{r},\textbf{p}){\rm d}^3 p\, {\rm d}^3 r = N$.

If the gas is trapped in a 3-D trap with a potential depth $U$, the collision integral in the Boltzmann equation can be evaluated analytically~\citep{Luiten96}. Indeed, the low-energy collisional cross-section
\begin{equation}
\sigma(k)=\frac{8\pi}{k^2+a^{-2}}
\end{equation}
reduces at unitarity to a simple dependence on the relative momentum of colliding partners: $\sigma(k)=8\pi/k^2$. However, not every collision leads to a loss of atoms due to evaporation. Consider
\begin{equation}
\eta=U/k_{\rm B} T \rm .
\label{Eq_eta}
\end{equation}
In the case of $\eta \gg 1$, such loss is associated with a transfer of large amount of energy to the atom which ultimately leads to the energy independent cross-section. This can be understood with a simple argument~\cite{Luo06}. Assume that two atoms collide with the initial momenta $\bm{p_1}$ and $\bm{p_2}$. After the collision they emerge with the momenta $\bm{p_3}$ and $\bm{p_4}$, and if one of them acquires a momentum $|\bm{p_3}| \gtrsim \sqrt{2 m U}$. Then, $|\bm{p_4}|$ is necessarily smaller than the most probable momentum of atoms in the gas and $|\bm{p_3}| \gg |\bm{p_4}|$. In the center of mass coordinates the absolute value of the relative momentum is preserved, so that $\frac{1}{2}|\bm{p_1}-\bm{p_2}|=\frac{1}{2}|\bm{p_3}-\bm{p_4}|\approx \frac{1}{2}|\bm{p_3}|$. Assuming $|\bm{p_3}| =\sqrt{2 m U}$, we get $|\bm{p_1}-\bm{p_2}|=\sqrt{2 m U}$. Substituting the relative momentum in the center of mass coordinate, $\hbar \bm{k}=\frac{1}{2}(\bm{p_1}-\bm{p_2})$, to the unitary form of the collisional cross-section, we find the latter is energy independent:
\begin{equation}
\sigma_{U}=\frac{16\pi\hbar^2}{mU},
\end{equation}
and the rate-equation for the atom number can be written as:
\begin{equation}
\frac{{\rm d}N}{{\rm d}t}=-\Gamma_{\rm ev}N, \;\;\;\; \Gamma_{\rm ev}=n_0 \sigma_U\overline{v}e^{-\eta}\frac{V_{\rm ev}}{V_{\rm e}}.
\label{Eq_GammaEv}
\end{equation}

The peak density is $n_0=N/V_{\rm e}$, where $V_e$ is the effective volume of the sample. In the harmonic trap $V_e$ can be related to $\overline{\omega}$ and the temperature $T$: $V_e=\left(\frac{2\pi k_{\rm B} T}{m\overline{\omega}^2} \right)^{3/2}$. The ratio of the evaporative and effective volumes is defined by~\cite{Luiten96}:
\begin{equation}
\frac{V_{\rm ev}}{V_{\rm e}}=\eta - 4 R\left(3,\eta\right),
\end{equation}
where $R(a,\eta)=\frac{P(a+1,\eta)}{P(a,\eta)}$ and $P(a,\eta)$ is the incomplete Gamma function

$$
P(a,\eta)=\frac{\int_0^\eta u^{a-1}e^{-u}{\rm d}u}{\int_0^{\infty}u^{a-1}e^{-u}{\rm d}u}.
$$

Finally, taking into account both three-body recombination loss (see Eqs.~(\ref{Eq_RateN3bSimple}),(\ref{Eq_Gamma3})) and evaporative loss, we can express the total atom number loss rate equation as:
\begin{equation}
\frac{{\rm d}N}{{\rm d}t}=-\gamma_3\frac{N^3}{T^5}-\gamma_2 e^{-\eta}\frac{V_{\rm ev}}{V_{\rm e}}\frac{N^2}{T},
\label{Eq_RateN}
\end{equation}
where

\begin{equation}
\gamma_2 = \frac{16}{\pi}\frac{\hbar^2 \overline{\omega}^3}{k_{\rm B} U}.
\end{equation}

Note that $\eta$ and the ratio of the evaporative and effective volumes explicitly depend on temperature and $\gamma_2$ is temperature independent.

\subsection{Rate equation for temperature}
\subsubsection{`Anti-evaporation' and recombination heating}
Ref.~\cite{Weber03} points out that in each three-body recombination event a loss of an atom is associated with an excess of $k_{\rm B}T$ of energy that remains in the sample.
This mechanism is caused by the fact that recombination events occur mainly at the center of the trap where the density of atoms is highest and it is known as `anti-evaporation' heating.
We now show that the unitary limit is more `anti-evaporative' than the regime of finite scattering lengths considered in ref.~\cite{Weber03} where $L_3$ is temperature independent.
We separate center of mass and relative motions of the colliding atoms and express the total loss of energy per three-body recombination event as following:

\begin{eqnarray}
\dot{E}_{\rm 3b}=-\int \left\lbrace \frac{L_3 n^3(\bm{r})}{3} \left(\langle E_{\rm cm} \rangle + 3U(\textbf{r})\right) \right. \nonumber\\
+ \left. \frac{n^3(\bm{r})}{3}\langle L_3(k)E_k \rangle \right\rbrace {\rm d}^3r.
\label{Eq_RateE3b}
\end{eqnarray}
The first two terms in the parenthesis represent the mean center-of-mass kinetic energy $\langle E_{\rm cm} \rangle = \langle P_{\rm cm}^2 \rangle /2M$ and the local potential energy $3U(\textbf{r})$ per each recombination triple.
$M=3m$ is the total mass of the three-body system.
The last term stands for thermal averaging of the three-body coefficient over the relative kinetic energy $E_k=(\hbar k)^2/2\mu$ where $\mu$ is the reduced mass.

Averaging the kinetic energy of the center of mass motion over the phase space density distribution (Eq.~(\ref{Eq_PhaseSpaceDist}))  gives $\langle E_{\rm cm} \rangle=\frac{3}{2} k_{\rm B}T$. Then the integration over this term is straightforward and using Eq.~(\ref{Eq_RateN3b}) we have:
\begin{equation}
-\int \frac{L_3 n^3(\textbf{r})}{3} \langle E_{\rm cm} \rangle {\rm d}^3 r =\frac{1}{2} k_{\rm B}T \dot{N}
\end{equation}

The integration over the second term can be easily evaluated as well:
\begin{equation}
-\int \frac{3 L_3}{3} n^3(\textbf{r}) U(\textbf{r}) {\rm d}^3r = \frac{1}{2}k_{\rm B} T \dot{N}
\end{equation}

To evaluate the third term we recall the averaged over the thermal distribution expression of the three-body recombination rate in Eq.~(\ref{Eq_FullL3}). Now its integrand has to be supplemented with the loss of the relative kinetic energy per recombination event $E_k$. Keeping the limit of Eq.~(\ref{Eq_ApproxL3}) this averaging can be easily evaluated to give $\langle L_3(k) E_k \rangle = L_3 k_{\rm B} T$. Finally, the last term in Eq.~(\ref{Eq_RateE3b}) gives:
\begin{equation}
- \int \frac{n^3(\textbf{r})}{3}\langle L_3(k)E_k\rangle {\rm d}^3r = \frac{1}{3}k_{\rm B} T \dot{N}
\end{equation}
Finally, getting together all the terms, the lost energy per lost atom in a three-body recombination event becomes:
\begin{equation}
\frac{\dot{E}_{\rm 3b}}{\dot{N}}=\frac{4}{3} k_{\rm B} T.
\label{Eq_LostEnergyPerLostAtom}
\end{equation}

This expression shows that unitarity limit is more `anti-evaporative' than the regime of finite scattering length ($k|a| \leq 1$). As the mean energy per atom in the harmonic trap is $3 k_{\rm B}T$, at unitarity each escaped atom leaves behind $(3 - 4/3)k_{\rm B} T =(5/3) k_{\rm B} T$ of the excess energy as compared to $1k_{\rm B} T$ when $L_3$ is energy independent. In the latter case, thermal averaging of the relative kinetic energy gives $\langle E_k \rangle = 3k_{\rm B} T$, thus $\dot{E}_{\rm 3b}/\dot{N} = 2k_{\rm B} T$.

Eq.~(\ref{Eq_LostEnergyPerLostAtom}) is readily transformed into the rate equation for the rise of temperature per lost atom using the fact that $E_{\rm 3b}=3 N k_{\rm B} T$ in the harmonic trap and Eq.~(\ref{Eq_RateN3bSimple}):
\begin{equation}
\frac{{\rm d}T}{{\rm d}t}=\frac{5}{3}\frac{T}{3} \gamma_3 \frac{N^2}{T^5}.
\label{Eq_RateT3b}
\end{equation}

Another heating mechanism pointed out in Ref.~\cite{Weber03} is associated with the creation of weakly bound dimers whose binding energy is smaller than the depth of the potential. In such a case, the three-body recombination products stay in the trap and the binding energy is converted into heat.

In the unitary limit, this mechanism causes no heating. In fact in this regime, as shown in the supplementary material to Ref.~\cite{Rem13}, the atoms and dimers are in chemical equilibrium with each other, e.g. the rate of dimer formation is equal to the dissociation rate. We therefore exclude this mechanism from our considerations.

\subsubsection{Evaporative cooling} \label{sec:model}
``Anti-evaporative'' heating can be compensated by evaporative cooling. The energy loss per evaporated atom is expressed as:
\begin{equation}
\dot{E}=\dot{N}\left(\eta  + \tilde{\kappa}\right)k_{\rm B} T
\label{Eq_Eve}
\end{equation}
where $\tilde{\kappa}$ in a harmonic trap is~\cite{Luiten96}:
\begin{equation}
\tilde{\kappa}=1-\frac{P\left(5,\eta\right)}{P\left(3,\eta\right)}\frac{V_{\rm e}}{V_{\rm ev}},
\end{equation}
with $0<\tilde{\kappa}<1$.

In a harmonic trap, the average energy per atom is $3k_{\rm B}T=\frac{E}{N}$. Taking the derivative of this equation and combining it with Eq.~(\ref{Eq_Eve}) we get:
\begin{equation}
3\frac{\dot{T}}{T}=\frac{\dot{N}}{N}\left(\eta + \tilde{\kappa}-3\right).
\label{Eq_Tev}
\end{equation}
From Eqs.~(\ref{Eq_GammaEv}) and (\ref{Eq_Tev}),  evaporative cooling is expressed as:
\begin{equation}
3\frac{{\rm d}T}{{\rm d}t}=-\Gamma_{\rm ev}\left( \eta + \tilde{\kappa} - 3\right) T,
\label{Eq_TevFF}
\end{equation}
This equation can be presented in a similar manner as in the previous section:
\begin{equation}
\frac{{\rm d}T}{{\rm d}t}=-\gamma_2 e^{-\eta}\frac{V_{\rm ev}}{V_{\rm e}} \left( \eta + \tilde{\kappa} - 3\right)\frac{N}{T}\frac{T}{3},
\label{Eq_Tevap}
\end{equation}
where, as before, the temperature dependence remains in $\eta$.

Finally, combining the two processes of recombination heating (Eq.~(\ref{Eq_RateT3b})) and evaporative cooling (Eq.~(\ref{Eq_Tevap})) we get:
\begin{equation}
\frac{{\rm d}T}{{\rm d}t}=\frac{T}{3}\left(\frac{5}{3}\gamma_3\frac{N^2}{T^5}-\gamma_2 e^{-\eta}\frac{V_{\rm ev}}{V_{\rm e}} \left( \eta + \tilde{\kappa} - 3\right)\frac{N}{T}\right).
\label{Eq_RateT}
\end{equation}

Eqs.~(\ref{Eq_RateN}) and (\ref{Eq_RateT}) form a set of coupled rate equations that describe the atom loss dynamics.

\subsection{N-T dynamics and the ``magic" $\eta_{\rm m}$}\label{ssNTdyn}

\begin{figure}
\centering\includegraphics[width=1.\columnwidth]{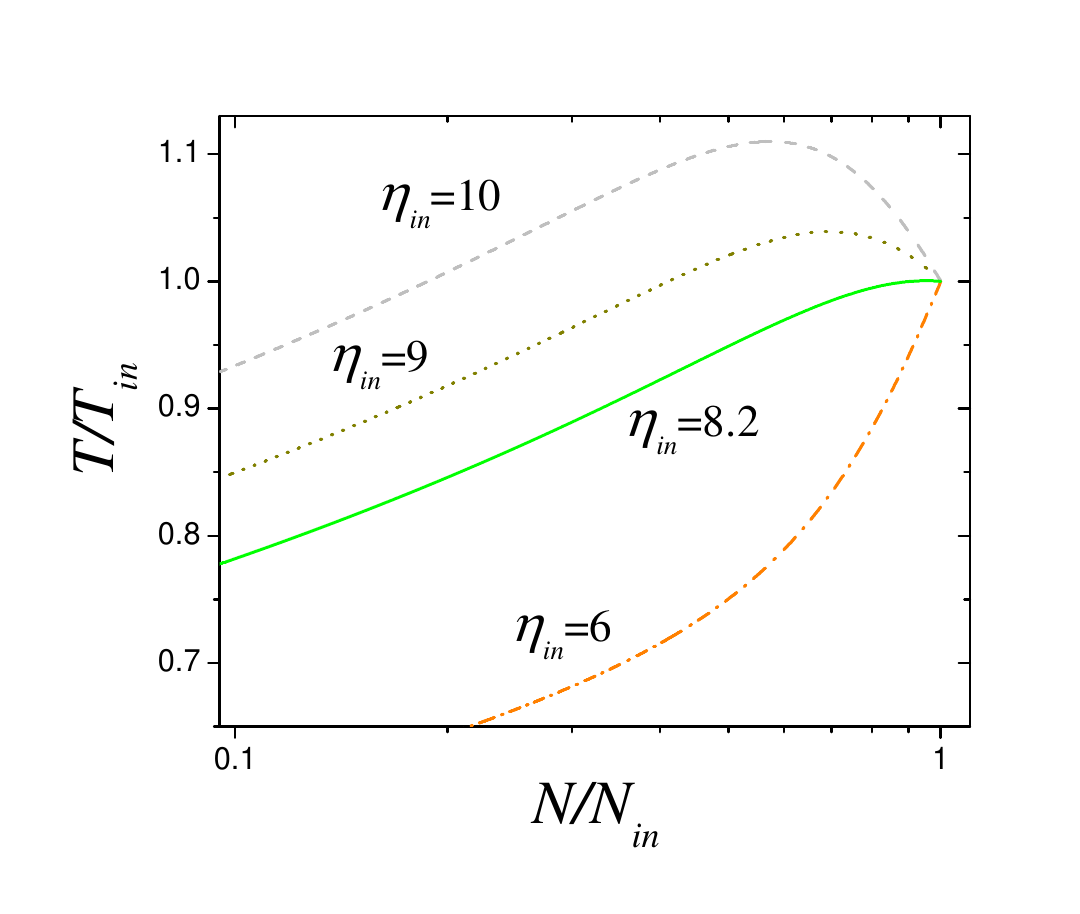}
\caption{\label{fig:NT_PhSp} N-T phase space representation of `anti-evaporation' heating and evaporative cooling dynamics for different values of the initial $\eta_{\rm in}$ parameter. The ``magic" $\eta_{\rm m}$ satisfies the condition ${\rm d}T/{\rm d}N=0$.  For lower values of $\eta_{\rm in}$ the ``magic" $\eta_{\rm m}$ is not reached during the evolution of the gas.The figure is drawn for experimental parameters of $^{133}$Cs atoms presented in Sec.~\ref{sec:Exp}. For these conditions $\eta_{\rm m}$ coincides with $\eta_{\rm in}\approx 8$.}
\end{figure}

To study atom number and temperature dynamics we solve Eqs.~(\ref{Eq_RateN}) and (\ref{Eq_RateT}) numerically for different initial values of $\eta$, referred to as $\eta_{\rm in}$ from here on.
As an illustration, $\gamma_2$ and $\gamma_3$ are evaluated based on parameters of the $^{133}$Cs experiment discussed in Sec.~\ref{sec:Exp}.
The system dynamics in $N-T$ phase space is represented in Fig.~\ref{fig:NT_PhSp}.
All represented values of $\eta_{\rm in}$ lead to a decrease in temperature for small atom numbers indicating that evaporative cooling always wins for asymptotic times where the atom density becomes small.
This weakens the three-body recombination event rate and effectively extinguishes the heating mechanism altogether.
Large values of the initial $\eta_{\rm in}$ cause initial heating of the system which is followed by a flattening of the temperature dependence at a certain atom number (grey dashed and dark yellow dotted lines) that defines the ``magic" $\eta_{\rm m}$.
In Fig.~\ref{fig:NT_PhSp} the solid green line represents the special case when $\eta_{\rm m}=\eta_{\rm in}$. Experimentally, $\eta(T,N)$ is tuned to satisfy this special case for a given initial temperature and atom number.
As it is seen in Fig.~\ref{fig:NT_PhSp}, lower initial values of $\eta_{\rm in}$ can never reach the necessary condition for $\eta_{\rm m}$ in their subsequent dynamics (orange dotted-dashed line).

\begin{figure}
\centering\includegraphics[width=1.\columnwidth]{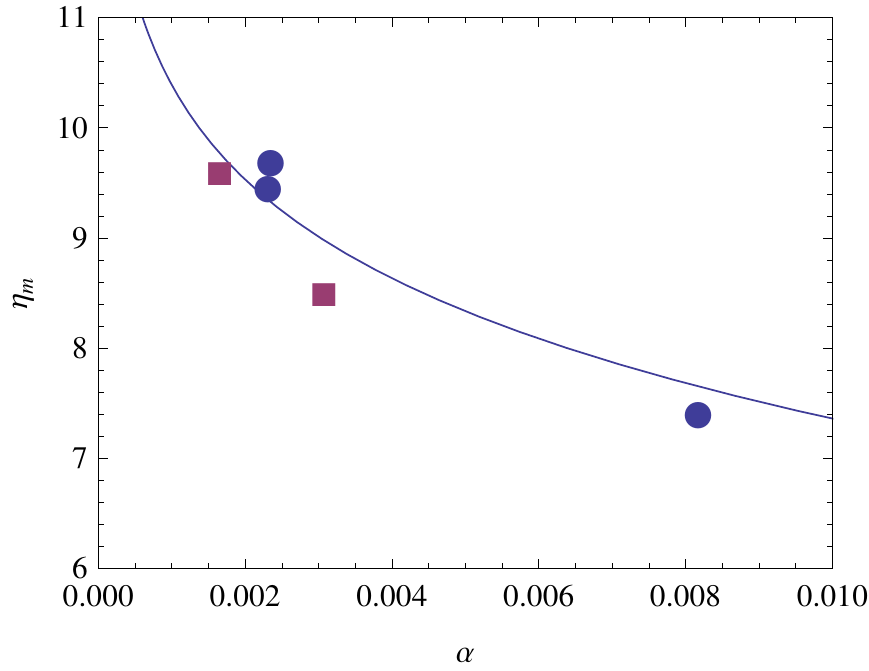}
\caption{\label{fig:scaling_magiceta} Universal plot $\eta_{\rm m}$ vs $\alpha=N(\hbar\bar\omega/k_B T)^3(1-e^{-4\eta_\ast})$ (bue curve). The blue solid circles correspond to the results obtained for $^{133}$Cs in Fig. \ref{FTvsNPlot2}(a) with $\eta_\ast= 0.098$. The red solid squares correspond to the $^7$Li data of Fig. \ref{FTvsNPlot2}(c) with $\eta_\ast=0.21$.}
\end{figure}

The value of $\eta_{\rm m}(T,N)$ is found by solving the equation ${\rm d}T/{\rm d}N=0$, i.e. when $T(N)$ becomes  independent on the atom number up to the first order in $N$. From the general structure of this equation, we see that $\eta_{\rm m}$ is solely function of the dimensionless parameter
\begin{equation}
\alpha=N\left(\frac{\hbar\bar\omega}{k_B T}\right)^3\left(1-e^{-4\eta_\ast}\right).
\label{Eq_alpha}
\end{equation}
Up to a factor $(1-e^{-4\eta_\ast})$, $\eta_{\rm m}$ depends only on the phase-space density $N(\hbar\bar\omega/k_B T)^3$ of the cloud. We plot in Fig.\,\ref{fig:scaling_magiceta} the dependence of $\eta_{\rm m}$ vs $\alpha$. Since our approach is valid only in the non-quantum degenerate regime where the momentum distribution is a Gaussian, we restricted the plot to small (and experimentally relevant) values of $\alpha$.

\subsection{2D evaporation}
\label{ss2D}
The above model was developed to explain 3D isothermal evaporation in a harmonic trap and experiments with $^{133}$Cs presented below correspond to this situation. Our model can also be extended to 2D isothermal evaporation, as realized in the $^7$Li gas studied in Ref.~\cite{Rem13} and presented below. In this setup, the atoms were trapped in a combined trap consisting of optical confinement in the radial direction and magnetic confinement in the axial direction. Evaporation was performed by lowering the laser beam power which did not lower the axial (essentially infinite) trap depth due to the magnetic confinement. Such a scenario realizes a 2D evaporation scheme.  Here, we explore the consequences of having 2D evaporation.  In the experimental section we will show the validity of these results with the evolution of a unitary $^{7}$Li gas.

Lower dimensional evaporation is, in general, less efficient than its 3D counterpart. 1D evaporation can be nearly totally solved analytically and it has been an intense subject of interest in the context of evaporative cooling of magnetically trapped hydrogen atoms~\cite{Luiten96,Surkov96,Pinkse98}. In contrast, analytically solving the 2D evaporation scheme is infeasible in practice.  It also poses a rather difficult questions considering ergodicity of motion in the trap~\cite{Mandonnet00}. The only practical way to treat 2D evaporation is Monte Carlo simulations which were performed in Ref.~\cite{Mandonnet00} to describe evaporation of an atomic beam. However, as noted in Ref.~\cite{Mandonnet00}, these simulations follow amazingly well a simple theoretical consideration which leaves the evaporation dynamics as in 3D but introduces an 'effective' $\eta$ parameter to take into account its 2D character.

The consideration is as following. In the 3D evaporation model, the cutting energy $\epsilon_{\rm c}$ is introduced in the Heaviside function that is multiplied with the classical phase-space distribution of Eq.~(\ref{Eq_PhaseSpaceDist})~\cite{Luiten96}. For the 2D scheme this Heaviside function is $Y(\epsilon_{\rm c}-\epsilon_\bot)$, where $\epsilon_{\rm c}$ is the 2D truncation energy and $\epsilon_\bot$ is the radial energy of atoms in the trap, the only direction in which atoms can escape. Now we simply add and subtract the axial energy of atoms in the trap and introduce an effective 3D truncation energy as following:
\begin{align}
Y(\epsilon_{\rm c}-\epsilon_\bot)=Y((\epsilon_{\rm c}+\epsilon_z)-(\epsilon_\bot+\epsilon_z))
=Y(\epsilon_{\rm c}^{\rm eff}-\epsilon_{\rm tot}),
\end{align}
where $\epsilon_{\rm tot}$ is the total energy of atoms in the trap and the effective truncation energy is given $\epsilon_{\rm c}^{\rm eff}=\epsilon_{\rm c}+\epsilon_z\simeq\epsilon_{\rm c}+k_{\rm B} T$  where we replaced $\epsilon_z$ by its mean value $k_{\rm B} T$ in a harmonic trap. The model then suggests that the evaporation dynamics follows the same functional form as the well established 3D model, but requires a modification of the evaporation parameter~\eqref{Eq_eta}: 
\begin{equation}
\eta^{\rm eff}=\eta + 1 \rm ,
\label{Eq_EffectiveEta}
\end{equation}
Then, the experimentally provided 2D $\eta$ should be compared with the theoretically found 3D $\eta^{\rm eff}$ reduced by 1 (i.e. $\eta^{\rm eff}-1$).

\section{Experiments}
\label{sec:Exp}

In this section, we present experimental $T(N)$ trajectories of unitary $^{133}$Cs and $^{7}$Li gases, and show that their dynamics are given by the coupled Eqns.\,\eqref{Eq_RateN} and \eqref{Eq_RateT}. The $^{133}$Cs Feshbach resonance at 47.8 Gauss and the $^7$Li Feshbach resonance at 737.8 Gauss have very similar resonance strength parameter $s_{\rm res}$ = 0.67 and 0.80 respectively~\cite{Lange2009,Chin10}) and are in the intermediate coupling regime (neither in the broad nor narrow resonance regime).
We first confirm the existence of a ``magic'' $\eta_{\rm m}$ for unitarity-limited losses for both species, with either 3D or 2D evaporation.
Then we will use the unitarity-limited three-body loss and the theory presented here to determine the Efimov inelasticity parameter of the narrow 47.8-G resonance in $^{133}$Cs which was not measured before.

\begin{figure}[]
\centering\includegraphics[width=1.\columnwidth]{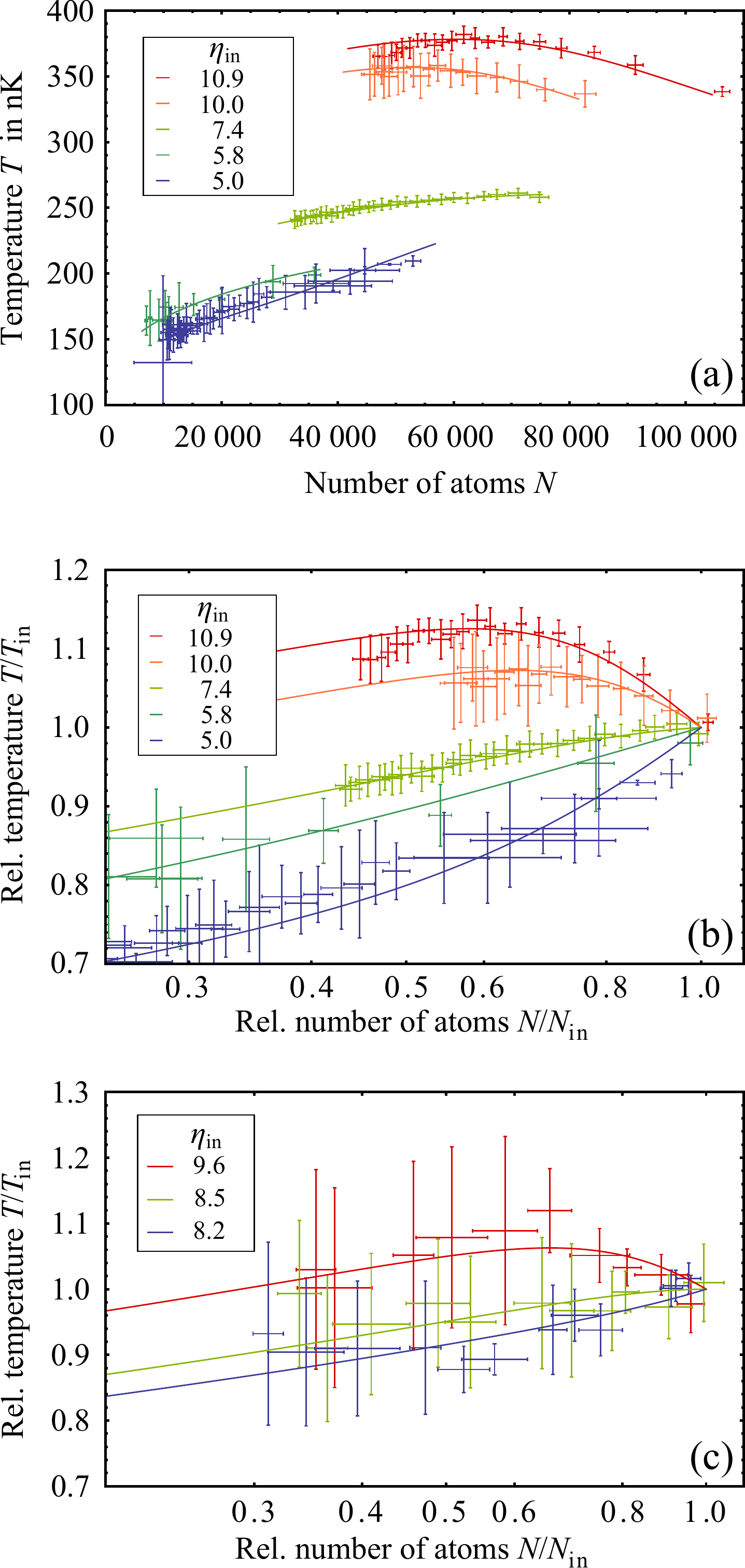}
\caption{(Color online) Evolution of the unitary $^{133}$Cs gas in (a) absolute and (b) relative numbers. The solid lines are fits of the data using the theory presented here, and the fitted initial relative trap depth $\eta_{\rm in} = U/k_{\rm B}T_{\rm in}$ is given in the legend. Error bars are statistical.  The  condition for $({\rm d}T/{\rm d}N)|_{t=0}$ is expected for $\eta_{\rm in}\approx \eta_{\rm m} \approx 8.2$, very close to the measured data for $\eta_{\rm in}= 7.4$ (green lines in a) and b)).
(c)~Evolution of the unitary $^{7}$Li gas. The solid lines are fits of the data using our 2D~ evaporation model, and the fitted initial relative trap depth $\eta_{\rm in} = U/k_{\rm B}T_{\rm in}$ is given in the legend. Error bars are statistical.  In 2D~evaporation, $\eta_{\rm in}\approx \eta_{\rm m}^{\rm eff} = \eta_{\rm m}+1=8.5$ is required to meet the $({\rm d}T/{\rm d}N)|_{t=0}$ condition, and is found in excellent agreement with the measured value 8.5 (green line in c)), see text.
}
\label{FTvsNPlot2}
\end{figure}

\subsection{$N-T$ fits}
\label{sssExpProc}

We prepare the initial samples at $T_{\rm in}$ and $N_{\rm in}$ as described in the Supplementary Materials. We measure the atom number $N(t)$ and the temperature $T(t)$ from \emph{in-situ} absorption images taken after a variable hold time $t$. In Fig.\,\ref{FTvsNPlot2}(a), we present typical results for the evolution $T(N)$ of the atom number and temperature of the gases, and we furthermore treat the hold time $t$ as a parameter. We also plot the relative temperature $T/T_{\rm in}$ as a function of the relative atom number $N/N_{\rm in}$ for the same data in Fig. \ref{FTvsNPlot2}(b), and for $^7$Li in Fig. \ref{FTvsNPlot2}(c). We then perform a coupled least-squares fit of the atom number and temperature trajectories, 
Eqs.\,\eqref{Eq_RateN} and \eqref{Eq_RateT}, to the data. We note that with our independent knowledge of the geometric mean of the trapping frequencies, $\bar{\omega}$, the only free fit parameters apart from initial temperature and atom number are the trap depth $U$ and the temperature-independent loss constant $\lambda_3$. 
 The solid lines are the fits (see Supplementary Materials) to our theory model, which describe the experimental data well for a large variety of initial temperatures, atom numbers and relative trap depth. We are able to experimentally realize the full predicted behavior of rising, falling and constant-to-first-order temperatures.

\subsection{Magic $\eta$}

The existence of maxima in the $T-N$ plots confirms the existence of a ``magic" relative trap depth $\eta_{\rm m}$, where the first-order time derivative of the sample temperature vanishes. 
Using the knowledge of $\eta_\ast$ for both $^{133}$Cs and $^7$Li, we can compare the observed values of $\eta_{\rm m}$ to the prediction of Fig. \ref{fig:scaling_magiceta} (note that in the case of $^7$Li, we plot $\eta_{\rm m}^{\rm eff}$ that enters into the effective 3D evaporation model). We see that for both the 3D evaporation $^{133}$Cs data and 2D evaporation  $^7$Li data, the agreement between experiment and theory is remarkable.

Furthermore, in the Supplemental Materials we show that from the three-body loss coefficients and the evaporation model, we can predict the trap depth, which is found in good agreement with the value deduced from the laser power, beam waist, and atom  polarizability.

\subsection{Universality of the three-body loss}

\begin{figure}[htb]
\centering\includegraphics[width=\columnwidth]{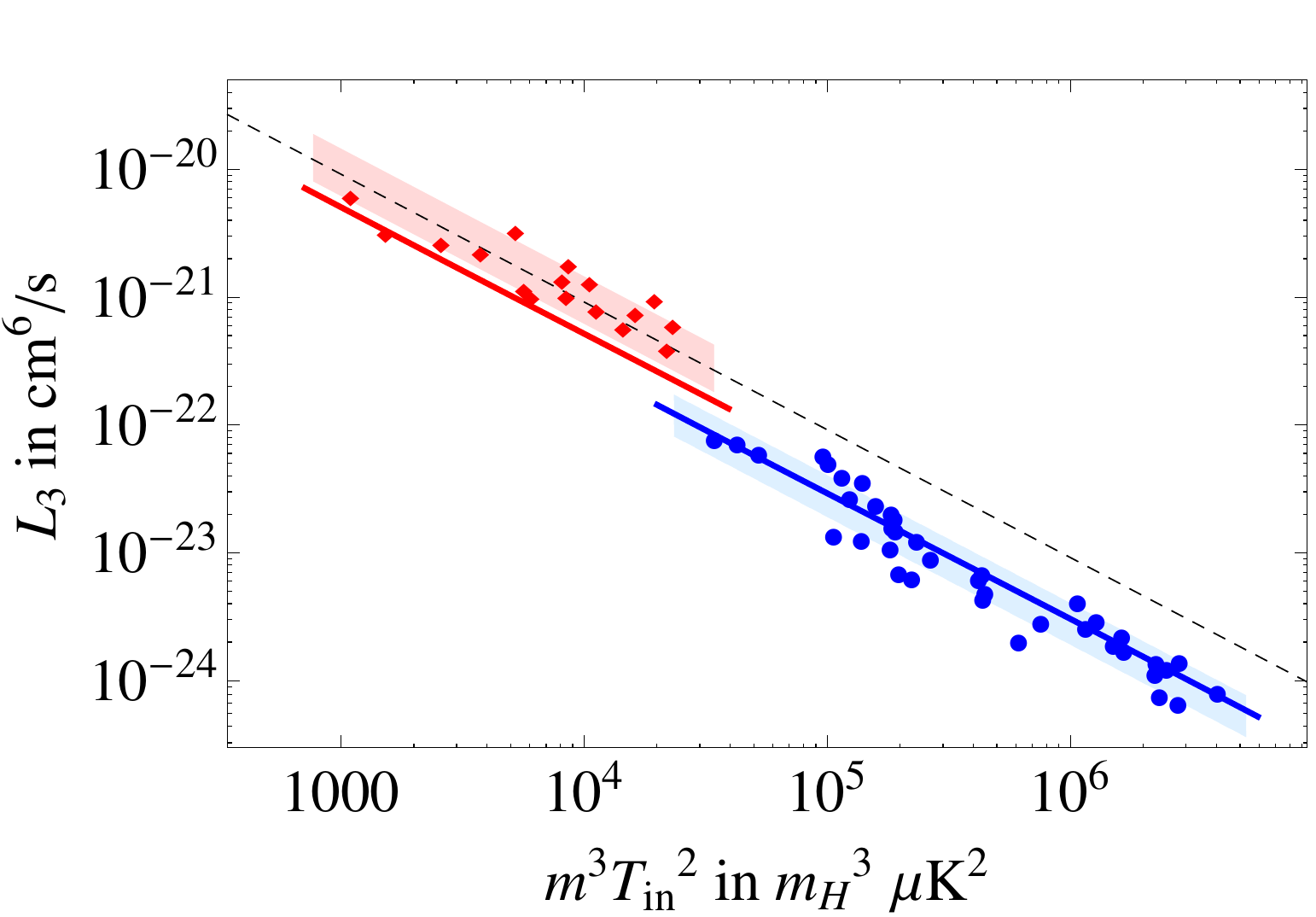}
\caption{(Color online) The magnitude of three-body loss rate at unitarity for $^7$Li (red) and $^{133}$Cs (blue)  with the respective $\pm 1$\, standard deviation (shaded areas). 
On the horizontal axis, masses are scaled to the hydrogen atom mass $m_{\rm H}$.
The dashed line represents the unitary limit (Eq.\,\eqref{Eq_ApproxL3} with $\eta_\ast \rightarrow \infty$). Solid lines are predictions of universal theory\,\cite{Rem13} with $\eta_\ast = 0.21$ for $^7$Li and $\eta_\ast = 0.098(7)$ for $^{133}$Cs, see text. The data confirms the universality of the $L_3 \propto T^{-2}$ law.
}
\label{FL3}
\end{figure}

 As the last application we now show the validity of the $L_3 \propto T^{-2}$ law for the tree-body loss of unitary $^7$Li and $^{133}$Cs Bose gases. Because both species are situated at the extreme ends of the (stable) alkaline group, they have a large mass ratio of 133/7 = 19 and the temperature range is varied over two orders of magnitude from 0.1\,$\upmu$K to 10\,$\upmu$K.
We determine the three-body loss coefficients $\lambda_3$ from fits to decay curves such as shown in Fig.\ref{FTvsNPlot2}. We present in Fig.\,\ref{FL3} the results for the rate coefficient $L_3$, which varies over approximately two orders of magnitude for both species.
In order to emphasize universality, the loss data is plotted as a function of $(m/m_{\rm H})^3 T_{\rm in}^2$, where $m_{\rm H}$ is the hydrogen mass. In this representation, the unitary limit for any species collapses to a single universal line (dotted line in Fig.\,\ref{FL3}, cf. Eq.\,\eqref{Eq_ApproxL3}).

For $^7$Li, we cover the 1-10$\,\upmu$K temperature range. We find for the temperature-independent loss coefficient $\lambda_3 = 3.0(3)\times 10^{-20}$\,cm$^6\upmu$K$^2$s$^{-1}$, very close to the unitary limit $\lambda_{3}^{\rm max} \approx 2.7\times 10^{-20}$\,cm$^6\upmu$K$^2$s$^{-1}$. It is also close to the value $\lambda_3 = 2.5(3)\times 10^{-20}$\,cm$^6\upmu$K$^2$s$^{-1}$ found in \cite{Rem13} with a restricted set of data, and to the predicion from Eq.\,\eqref{Eq_FullL3} with $\eta_\ast = 0.21$ from \cite{Gross10} (red solid line in Fig.\,\ref{FL3}).
We cannot measure $\eta_\ast$ here because the $^7$Li data coincides with the unitary limit. 

Furthermore the quality of the $^{133}$Cs temperature and atom number data enables us to directly measure the previously unknown $\eta_\ast$ parameter of the 47.8-G Feshbach resonance.
The standard technique for obtaining $\eta_\ast$ is measuring the three-body loss rate $L_3(a,T\rightarrow 0)$ as a function of scattering length in the zero-temperature limit, and subsequent fitting of the resulting spectrum to universal theory. However, for a given experimental magnetic field stability, this method becomes hard to put into practice for narrow resonances like the 47.8-G resonance in $^{133}$Cs.
Instead, we use the fits to our theory model in order to obtain $\eta_\ast$ from $\lambda_3$.
We cover the 0.1-1\,$\upmu$K range and find $\lambda_3 = 1.27(7)\times 10^{-24}$\,cm$^6\upmu$K$^2$s$^{-1}$.
Plugging this number into Eq.\,\eqref{Eq_ApproxL3}, we deduce a value for the Efimov inelasticity parameter
$\eta_\ast = 0.098(7)$. 
The corresponding curve is the blue line in Fig.\,\ref{FL3} and is significantly below the unitary line because of the smallness of $\eta_\ast$.
This new value is comparable to the Efimov inelasticity parameter found for other resonances in $^{133}$Cs, in the range 0.06...0.19\,\cite{Kraemer06,Berninger11}.

The plot of the full theoretical expression Eq.\,\eqref{Eq_FullL3} for $L_3(m^3T^2)$ in Fig.~\ref{FL3} (full lines) requires an additional parameter describing three-body scattering around this Feshbach resonance, the so-called three-body parameter. It can be represented by the location of the first Efimov resonance position $a^{(1)}_{-}$~\cite{Chin2011}. Because of the lack of experimental knowledge for the 47.8-G resonance, we take the quasi-universal value {$a^{(1)}_{-} = - 9.73(3) r_{\rm vdW}$}, $r_{\rm vdW}$ being the van-der-Waals radius, for which theoretical explanations have been given recently\,\cite{Chin2011,Wang2012a,Naidon2014}.
The theory curve then displays log-periodic oscillations with a temperature period set by the Efimov state energy spacing of $\exp(2\pi/s_0) \approx 515$, where $s_0= 1.00624$, and with a phase given by $a^{(1)}_{-}$. The relative peak-to-peak amplitude is 7\% for $^{133}$Cs. As seen in Fig.\,\ref{FL3}, such oscillations cannot be resolved in the experimental data because of  limited signal-to-noise and the limited range of temperature. The predicted contrast of these oscillations for $^7$Li is even smaller ($\sim 6\%$). This is a general property of the system of three identical bosons due to the smallness of $|s_{11}|$~\cite{Rem13}.

\section{Conclusions}

In this article, we developed a general theoretical model for the coupled time dynamics of atom number and temperature of the 3D harmonically trapped unitary Bose gas in the non-degenerate regime. The theory takes full account of evaporative loss and the related cooling mechanism, as well as of the universal  three-body loss and heating. It is furthermore extended to the special case of 2D evaporation. We predict  and experimentally verify the existence of a ``magic'' trap depth, where the time derivative of temperature vanishes both in 3D and 2D evaporation.

We compare our model to two different set of experiments with lithium and cesium with vastly different mass and temperature ranges. The data illustrates the universal $T^{-2}$ scaling over 2 orders of magnitude in temperature, and we obtain an experimental value of the Efimov inelasticity parameter for the 47.8-G resonance in $^{133}$Cs.
The theory further enables an independent determination of the trap depth. The  agreement found here with standard methods  shows that it can be used in more complex trap geometries (crossed dipole traps, or hybrid magnetic-optical traps) where the actual trap depth is often not easy to measure.

In future work it would be highly interesting to probe the discrete symmetry of the unitary Bose gas by revealing the 7\% log-periodic modulation of the three-body loss coefficient expected over a factor 515 energy range.

\acknowledgments{We would like to thank the Institut de France (Louis D. award), the region Ile de France DIM nanoK/IFRAF (ATOMIX project), and the European Research Council ERC (ThermoDynaMix grant) for support. 
We acknowledge support from the NSF-MRSEC program, NSF Grant No. PHY-1206095, and Army Research Office Multidisciplinary University Research Initiative (ARO-MURI) Grant No. W911NF-14-1-0003.
L.-C. H. is supported by the Grainger Fellowship and the Taiwan Government Scholarship. We also acknowledge the support from the France-Chicago Center.}

\bibliography{FewBody}
\bibliographystyle{unsrt}

\widetext
\newpage
\begin{center}
\textbf{\large Supplemental Material: Universal Loss Dynamics in a Unitary Bose Gas}
\end{center}
\setcounter{equation}{0}
\setcounter{figure}{0}
\setcounter{table}{0}
\setcounter{page}{1}
\makeatletter
\renewcommand{\theequation}{S\arabic{equation}}
\renewcommand{\thefigure}{S\arabic{figure}}
\renewcommand{\thepage}{S\arabic{page}}

\begin{figure*}[b]
\centering\includegraphics[width=\textwidth]{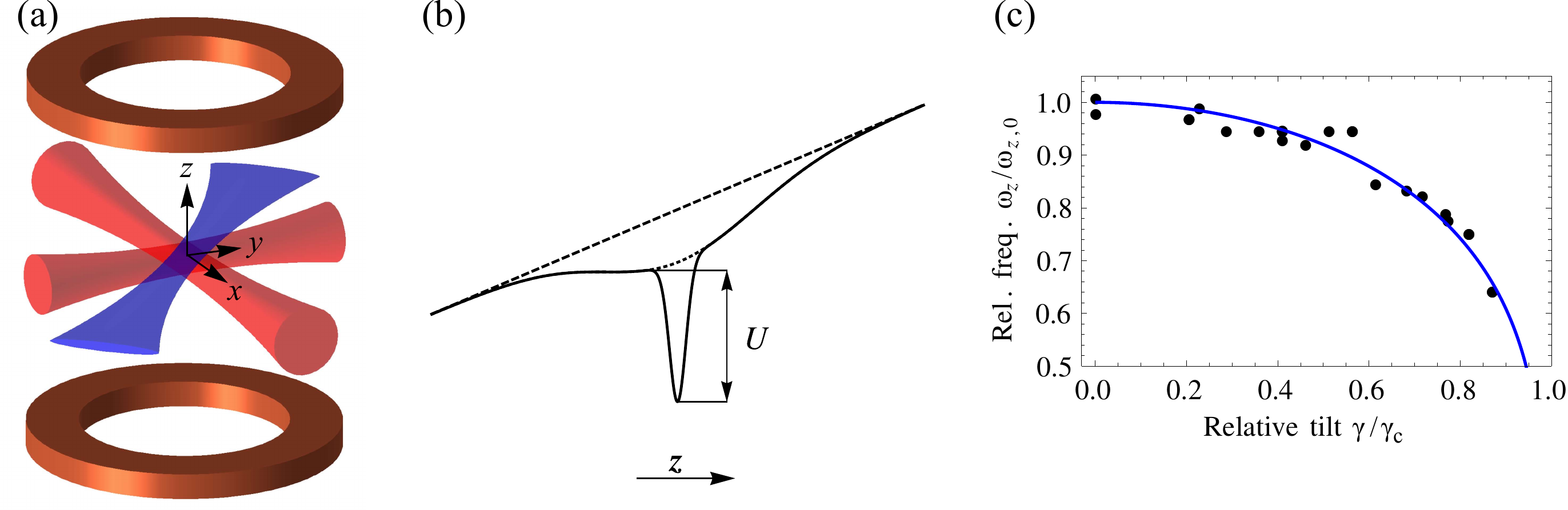}
\caption{(Color online) (a) Schematic drawing of the hybrid trap. It consists of three intersecting lasers, and a magnetic field gradient in the $z$\,direction (vertical) created by a pair of coils. The light sheet beam (blue) confines dominantly along the vertical direction. The additional round beams (red) stabilize the horizontal confinement. (b) Trap shape along the vertical ($z-$) direction. It is composed of two Gaussians and a linear contribution from the tilt, see Eq.\,\eqref{EVz}. Relative dimensions are to scale. Also indicated are the contributions of the tilt only (dashed line) and the round beams (dotted line). (c) 
Trap frequency measurements (dots) and fits (line) as a function of the tilt of the trap. The frequencies are normalized with respect to the zero-tilt frequency $\omega_{z,0}= 2\pi\times 140$\,Hz. We normalize the tilt from our knowledge of the critical $\gamma_{\rm c}$, where the trap opens, see text.} 
\label{FTrap}
\end{figure*}

\subsection{$^{133}$Cs setup}

Our setup is a modified version of the one presented in\,\cite{Hung08}. The $^{133}$Cs atoms are trapped by means of three intersecting laser beams, and a variable magnetic field gradient in the vertical direction (partially) compensates gravity. An intrinsic  advantage of the scheme is the perfect spin polarization in the lowest hyperfine ground state $|F,m_F \rangle = |3,3\rangle$, because the dipole trap potential is too weak to hold atoms against gravity if they are in any other ground state. As we will see, the trap frequencies stay almost constant when reducing the trap depth, making evaporation very efficient\,\cite{Hung08}.

\subsection*{Trap model}
\label{ssTrapModel}

The trap consists of three 1064-nm laser beams and an additional magnetic gradient field, see Fig.\,\ref{FTrap}\,(a). All beams propagate in the horizontal plane. An elliptical light sheet beam (power $P_{\rm LS} = $520\,mW, waists: $w_{\rm LSv} = 33.0$\,$\upmu$m [vertical direction] and $w_{\rm LSh} = 225\,\upmu$m [horizontal direction]) creates the vertical confinement together with the magnetic gradient. Two round beams ($P_{\rm R1,R2} = $1.1/1.2\,W, waist of 
$w_{\rm R} = 300\,\upmu$m
) stabilize the horizontal confinement. The light sheet center is $z_0 = 6\,\upmu$m lower than the center of the trap formed by the round beams only. The potential along the vertical axis can therefore be written as
\begin{equation}
V(z) = -U_{\rm LS}\,{\rm e}^{-\frac{2 z^2}{w_{\rm LSv}^2}} - U_{\rm R}\,{\rm e}^{-\frac{ 2 (z - z_0)^2}{w_{\rm R}^2}} + \gamma z
\rm ,
\label{EVz}
\end{equation}
where the $U_{\rm LS}$ and $U_{\rm R}$ are the contributions from the light sheet and round beams, respectively. 
The tilt $\gamma$ has a gravitational and a magnetic contribution,
\begin{equation}
\gamma = m g - \mu B'
\rm ,
\label{EGamma}
\end{equation}
where $m$ is the 
 atomic mass, $g$ is the gravitational acceleration, $B' = \partial_z B_z$  
is the magnetic field gradient along the $z$-axis, and $\mu = 0.75\,\mu_{\rm B}$ is the atom's magnetic moment in the $|3,3\rangle$ state, with $\mu_{\rm B}$ being Bohr's magnetic moment. Thus, a gradient of $B'_0 = B'(\gamma = 0) = 31.3\,$G/cm is needed for magnetically levitating the cloud. An example potential shape is given in Fig.\,\ref{FTrap}\,(b).

\subsection*{Trap frequency calibration}

When we intentionally change the trap depth, we also change the trap frequencies, mostly affecting the vertical direction. The data was taken during two different measurement campaigns in 2012 and 2013. Therefore, the trap had to be recalibrated for each of this campaigns, and the data is presented in a normalized way.

We measure the oscillation frequency $\omega_z$ along the $z$\,axis as a function of the tilt, see Fig.\,\ref{FTrap}\,(c). This is established by inducing sloshing oscillations to a small, weakly-interacting Bose-Einstein condensate (BEC), and performing time-of-flight measurements of its position after a variable hold time.

We fit the measured $z$-axis frequencies to a numerical model of the trap potential Eq.\,\ref{EVz}. In the model, we plug the aspect ratio of the trap depth contributions from the three beams $i$,
\begin{equation}
U_i = \frac{2}{\pi}\frac{\alpha P_i}{w_{\rm h} w_{\rm v}}
\rm ,
\label{EU}
\end{equation}
where $\alpha/k_{\rm B} =  2.589 \times 10^{-12} {\rm K.}{\rm cm^{2}.mW^{-1}}$ is the atomic polarizability at 1064\,nm\,\cite{Hung2013}, $P_i$ is the power in beam $i$, and the $w_{\rm h/v}$ are the waists in the horizontal or vertical direction. We are left with two fit parameters:
The frequency at zero tilt $\omega_z(\gamma = 0) = \omega_{z,0}$, and the critical tilt $\gamma_{\rm c} = \gamma(B' = B'_{\rm c})$ where the trap opens (local minimum in $V(z)$ disappears) and $\omega_z$ goes to zero by construction.  We find $\omega_{z,0} = 2\pi\times 139(1)$\,Hz and $2\pi\times140(1)$\,Hz for 2012 and 2013, and
{$B'_{\rm c} = -0.3(4)$\,G/cm and $-4.3(7)$\,G/cm} for 2012 and 2013.
With this calibration, we introduce the normalized tilt $\gamma/\gamma_{\rm c}$.
The values for $B'_{\rm c}$ coincide well with the gradient values observed when increasing the tilt until a small ($<~5000$ atoms) weakly-interacting BEC drops out of the trap.

The kink in the trap depth theory curve (Fig.\,\ref{FUCsLi}) near {$\gamma/\gamma_{\rm c} = 0.02$} (shaded area in Fig.\,\ref{FUCsLi}(a)) corresponds to a situation depicted in Fig.\,\ref{FTrap}\,(b), where the contribution of the large-waist horizontal beams on the trap depth vanishes. 
The blue-shaded region of the horizontal beams' contribution extends over the small region from {$\gamma/\gamma_{\rm c} = -0.02$} to 0.02. Therefore, small experimental uncertainties on the applied magnetic gradient, or additional trap imperfections can explain the fact that we do not find this sudden rise in $U$. Other than that, we see a remarkable correspondence between theory and experiment.

We note that the data can also be well described by the analytical model of a single gaussian potential ($U_{\rm R} = 0$) with tilt, as presented in\,\cite{Hung08}. 
Because of the large mismatch between $w_{\rm R}$ and $w_{\rm LS}$, the presence of the round beams mainly affects the horizontal trapping. The critical gradient we find is only 2\% larger than the single-gaussian value $\sqrt{{\rm e}} \gamma w_{\rm LS}/ 2 U_{\rm LSv}$\,\cite{Hung08}. Furthermore, the horizontal trapping frequencies $(\omega_{x},\omega_y) \approx 2\pi\times(13,30)$\,Hz, measured with a similar method for each dataset, remain constant.

\subsection*{Imaging system calibration}

The high-resolution imaging system is similar to the one presented in\,\cite{Ha2013}. It is well calibrated using the equation of state of a weakly-interacting 2D Bose gas for the absorption-coefficient-to-atomic-density conversion (in good accordance with the method of classical 2D gas atomic shot noise\,\cite{Hung2011b}). The imaging magnification is obtained from performing Bragg spectroscopy on a 3D BEC, using the variable retroreflection of the 1064-nm round beams.

\subsection*{$^{133}$Cs sample preparation}

We prepare the $^{133}$Cs samples in the trap described before. In brief, after magneto-optical trapping and degenerate Raman sideband cooling we obtain magnetically levitated ($\gamma = 0$) samples of $10^6$ $^{133}$Cs atoms at $1\,\upmu$K\,\cite{Hung08}. We can cool the samples further by evaporative cooling. In order to achieve this, we adjust the trap depth $U(\gamma)$ by changing the tilt $\gamma$ of the potential \eqref{EVz}.
Thus, the samples can be evaporatively cooled all the way to quantum degeneracy in $\approx$\,2\,s\,\cite{Hung08} at {20.8\,G, yielding a scattering length of 200}\,$a_0$, with $a_0$ being the Bohr radius\,\cite{Kraemer06}.

We prepare our samples by stopping the evaporation at a given tilt. We then ramp the tilt adiabatically to the desired value. Finally, at a time $t_0$, we jump the field to the Feshbach resonance at 47.8\,G\,\cite{Lange2009} in typically $<~1$\,ms and wait for at least $2\pi/\omega_x$ in  order for the samples to reach dynamical equilibrium. We are therefore able to prepare samples of variable initial parameters: Atom number $N_{\rm in}$, temperature $T_{\rm in}$ and relative trap depth $\eta_{\rm in} = U(\gamma)/k_{\rm B}T_{\rm in}$, where $k_{\rm B}$ is the Boltzmann constant. After a hold time $t$, we take an \emph{in-situ} absorption image with a vertical imaging setup.

\subsection{$^{7}$Li setup}

The $^{7}$Li data was taken using the apparatus described in~\cite{Rem13}.  This trap consists of a 1073-nm single-beam optical dipole trap providing adjustable radial confinement, and an additional magnetic field curvature providing essentially infinitely deep harmonic axial confinement along the beam axis.   After loading into this trap, the gas is evaporated by lowering the radial trap depth at a magnetic field of 720\,G, where the 2-body scattering length is 200\,$a_{0}$.  The evaporation in this hybrid trap is then effectively 2D. After the temperature and atom number of the gas have stabilized, the radial trap is adiabatically recompressed by 
about a factor of two.  Since the 
axial magnetic confinement is practically unchanged, this recompression causes the temperature of the gas to increase with $\omega_{r}^{3/2}$, while the trap depth increases as $\omega^2$.  Consequently, by varying the amount of recompression, we can vary $\eta_{\rm in}$.  After this recompression, the magnetic field is ramped to the Feshbach resonance field of 737.8~G in 100-500~ms and $N(t)$ and $T(t)$ are measured with in situ resonant absorption imaging perpendicular to the long axis of the cloud.
The trap shape can be described by Eq.\,\eqref{EVz} with $U_{\rm LS} = 0$, $w_{\rm R} = 37(1)\,\upmu$m, and replacing $z$ by $\rho$, the radial coordinate. $U_{\rm R}$ is the  dipole trap potential with power $P_{\rm trap}$ and $\alpha$ is the polarizability of the $^7$Li atoms at 1073\,nm. We can neglect the tilt $\gamma$ because of the small mass of $^7$Li.

\subsection{Time scale order}

In order for the theory to be valid, we make sure the timescale order is not violated:
\begin{equation}
\tau_{\rm 3B}, \tau_{\rm ev} \gg \tau_{\rm trap}, \tau_{\rm 2B}
\rm ,
\end{equation}
where we have the three-body loss time constant (cf. Eq.\,(19)
)
\begin{equation}
\tau_{\rm 3B}^{-1} = \frac{5}{9} \gamma_3 \frac{N^2}{T^5}
\rm ,
\end{equation}
the evaporation time constant (cf. Eq.\,(25)
)
\begin{equation}
\tau_{\rm ev}^{-1} = \frac{1}{3}\gamma_2 \left(e^{-\eta}\frac{V_{\rm ev}}{V_{\rm e}}\right) \left( \eta + \tilde{\kappa} - 3\right)\frac{N}{T}
\rm ,
\end{equation}
the two-body scattering time constant
\begin{equation}
\tau_{\rm 2B}^{-1} = n_0 \sigma_U \bar{v}
\rm ,
\end{equation}
and the trapping time constant
\begin{equation}
\tau_{\rm trap}^{-1} = \omega_{\rm slow}
\rm ,
\end{equation}
where $\omega_{\rm slow}$ is the slowest trapping frequency (along $z$ in the $^{133}$Cs case).

\subsection{Fits to the model}

For each data set, we have decay data for $N(t)$ and $T(t)$. We fit both temperature and atom number individually with solutions to the coupled differential equation set of Eqs.\,\eqref{Eq_RateN} and \eqref{Eq_RateT}. For both fits, we use a common three-body loss coefficient $\lambda_3$, and a common trap depth $U$. The fitting is done by minimizing the weighted sum $\alpha \chi_T + \alpha^{-1} \chi_N$ by varying both the weighing factor $\alpha$ and the fit parameters. The quadratic deviations are defined as  $\chi_{T,N} = \Sigma \sigma_{T,N}^2$ ($\sigma_{T,N}$ being the deviations of data and fit). This method also accounts for the different amount of relative signal-to-noise ratio of both data sets.

\subsection{Trap depth}

\begin{figure*}[]
\centering\includegraphics[width=\textwidth]{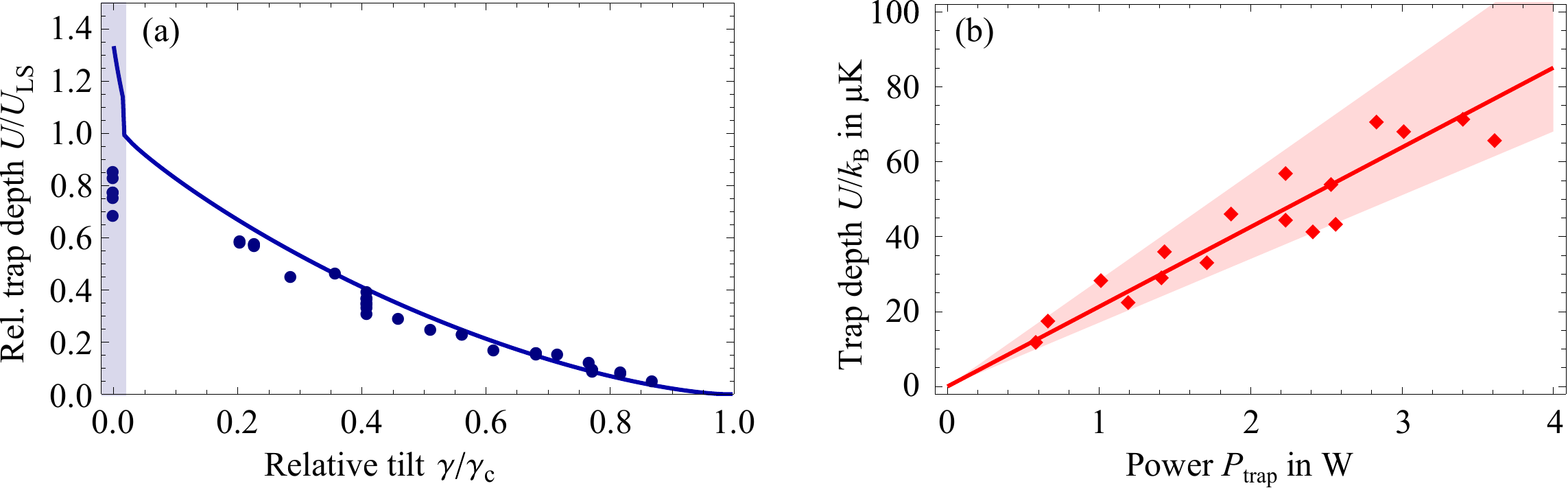}
\caption{(Color online) {(a)}
Relative $^{133}$Cs trap depth results from the fits to our data (dots), and theoretical model (Eqs.\,\eqref{EVz} and \eqref{EU}, line). The trap depth is normalized with respect to the trap depth $U_{\rm LS} \approx 11\,\upmu$K given by the gaussian light sheet only. The shaded area corresponds to the region where the horizontal confinement beams significantly contribute to the trap depth.
{(b)}
Absolute $^7$Li trap depth results from the fits to our data (dots). The solid line indicatess theoretical knowledge of our trap (Eqs.\,\eqref{EVz} and \eqref{EU}, line), with $w_{\rm R} = 38(1)\,\upmu$m. The shaded area accounts for the combined uncertainty of $w_{\rm R}$ and $P_{\rm trap}$.
}
\label{FUCsLi}
\end{figure*}

As an independent test of the theory fits, we compare the fitted trap depth $U$ to its independently known counterpart from experimental parameters. 
In Fig.\,\ref{FUCsLi}(a), we plot the $^{133}$Cs results as a function of the relative trap tilt $\gamma/\gamma_{\rm c}$. 
We also plot the theoretical value for $U(\gamma/\gamma_{\rm c})$ as a solid line. 
Except near zero tilt, we find excellent agreement of the fitted values with the values known from experimental parameters. 
For $^7$Li, see Fig.\,\ref{FUCsLi}(b), we find excellent agreement with our theoretical knowledge of the trap depth, which is given by the dipole laser waist $w_{\rm R}$, power $P_{\rm trap}$ and the atom's polarizability. It is indicated by the shaded area in  Fig.\,\ref{FUCsLi}(b). Therefore, we can infer the dipole trap laser's waist in an independent fashion. From the fit to our measured trap depths (solid line) in Fig.\,\ref{FUCsLi}(b) we obtain $w_{\rm R} = 38(1)\,\upmu$m. This value coincides with independent measurements of $w_{\rm R}$ from fitting the trap frequencies as a function of $P_{\rm trap}$. These results emphasize the validity of the theory model (Eqs.\,\eqref{Eq_RateN} and \eqref{Eq_RateT}).


\end{document}